# Reconfiguring Geovisualization in the Age of Generative AI: Insights from Domain Experts

Mengyi Wei[a], Chenyu Zuo[b,c], Jiaying Xue[a], Nianhua Liu[a], Dongsheng Chen[a], Shengkai Wang[a], Yu Feng[a]*, and Liqiu Meng[a]

[a]*Chair of Cartography and Visual Analytics, Technical University of Munich, Germany;* [b]*Geoinformatics Lab, Institute of Geography, University of Augsburg, Germany;* [c]*Department of Civil, Environmental and Geomatic Engineering, ETH Zürich, Switzerland*

Generative artificial intelligence (GenAI) is increasingly integrated into geovisualization, yet its broader implications for professional practice are insufficiently understood. To examine these implications, we conducted semi-structured interviews with 20 geovisualization experts from Europe, the United States, and China. The interviews were structured around four broad analytical domains: Data, Ideation, Prototyping, and Iteration, while also encouraging participants to reflect on issues that extend beyond these activities.

Our findings show that GenAI expands the capabilities of geovisualization, particularly in terms of data handling, creative exploration, and rapid prototyping, but does not simply remove existing constraints. Instead, key bottlenecks are shifting from production to judgment and verification. As routine technical tasks become more automated, professional value increasingly depends on spatial reasoning, contextual interpretation, aesthetic and ethical judgment, and the ability to assess whether AI-generated outputs are appropriate for use. At the same time, GenAI introduces new challenges regarding provenance, interpretability, and accountability, raising questions about how responsibility should be distributed across models, developers, practitioners, institutions, and users.

These shifts are particularly significant in geovisualization because spatial representations are constrained by geographic reality and must balance scientific validity, visual expression, and technical implementation. We therefore argue that responsible GenAI in geovisualization requires domain-specific approaches to spatial validation, provenance, uncertainty communication, human oversight, and accountable use. This study provides an expert-grounded perspective on how GenAI is reconfiguring geovisualization as a practice of spatial knowledge production. It also identifies implications for future professional practice, education, system design, and governance.

## 1. Introduction

Breakthroughs in Generative AI (GenAI), such as GPT-4, Claude, Stable Diffusion, DALL·E, and Midjourney, are reshaping many professional domains centered on knowledge production and expression (Sowa & Przegalinska, 2025). From natural language processing (Vaswani et al., 2017) and text summarization (Lewis et al., 2019) to image generation (Rombach et al., 2022) and multimodal content creation (Ramesh et al., 2022), the capabilities of GenAI are rapidly expanding. These developments have attracted growing attention not only in traditionally creative domains but also in geovisualization, where scientific data analysis, visual expression, and technological implementation are combined to support scientific inquiry and complex decision-making(G. Andrienko et al., 2014; Çöltekin et al., 2018).

The impact of GenAI on geovisualization extends beyond technical innovation. It reflects a broader transformation within a socio-technical field that integrates data, knowledge, interpretation, and expression (MacEachren & Kraak, 2001). Research on these implications remains at an early stage, while technical applications, stylistic innovations, and ethical concerns continue to evolve rapidly. Literature-based studies alone may not fully capture emerging practices and changing disciplinary perspectives. We therefore conducted semi-structured interviews with geovisualization experts, whose domain knowledge, practical experience, and forward-looking perspectives can help identify underexplored developments in the field (Bogner et al., 2009).

To investigate these issues, this study uses four broad activity domains to structure the inquiry: *Data, Ideation, Prototyping, and Iteration* (Basole & Major, 2024; Çöltekin et al., 2018; Kraak, 2003; Smith et al., 2013). These domains are not proposed as an exhaustive, universal, or sequential model of geovisualization. Rather, they serve as an analytical scaffold for eliciting and comparing expert perspectives on recurring forms of work that are currently or potentially affected by GenAI. The four-domain structure provides sufficient analytical coverage while maintaining the openness required for semi-structured expert interviews (Figure 1).

Building on this structure, we first examine how experts perceive current challenges in geovisualization. We then investigate their views on the applications and impacts of GenAI, including emerging forms of human–AI collaboration and implications for professional practice. Finally, we explore the risks associated with GenAI and possible approaches to addressing them. The study is guided by the following research questions:

**RQ1:** How do geovisualization experts perceive the current challenges facing the field?

**RQ2:** What impact does the emergence of GenAI have on geovisualization?

**RQ3**: How do experts in the field assess and respond to the potential risks posed by GenAI?

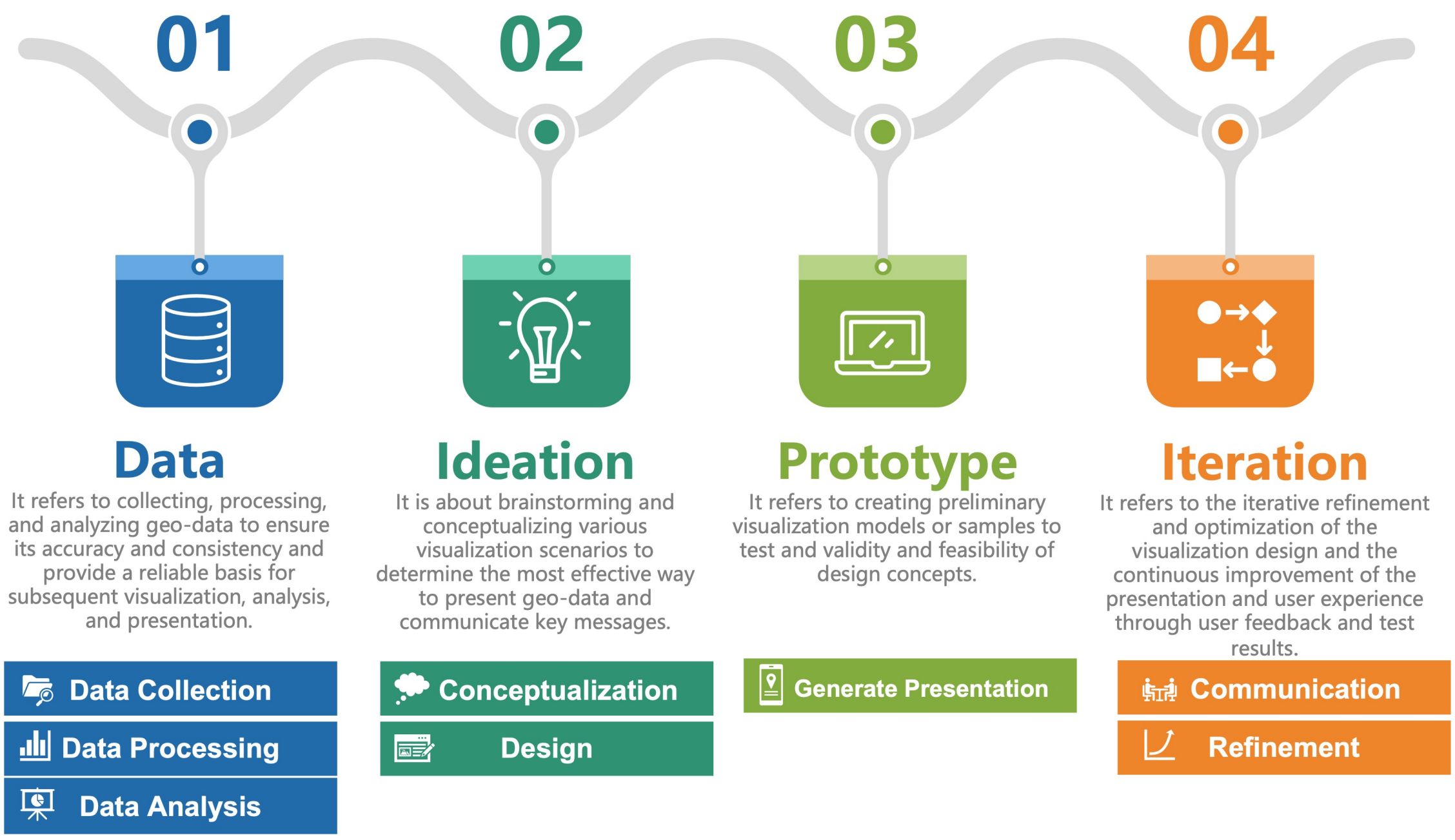

Figure 1. Four broad activity domains used to structure the investigation of GenAI in geovisualization.

To address these questions, we interviewed 20 geovisualization experts affiliated with universities and research institutions in Europe, the United States, and China. Rather than treating GenAI primarily as a new production tool, this study examines how its adoption may redistribute bottlenecks, professional expertise, and responsibility across geovisualization practice. By drawing on socio-technical and human–AI interaction perspectives, we provide an expert-grounded account of how GenAI is reconfiguring geovisualization and identify implications for future professional practice, education, system design, and governance.

## 2. Related Work

This section reviews three strands of research relevant to the broader implications of GenAI for geovisualization: the scientific and human-centered development of geovisualization, human–AI collaboration in creative practice, and emerging GenAI applications in geospatial data processing and visualization. Together, these strands establish the disciplinary, collaborative, and technological foundations of the present study.

### *2.1 Scientific Development*

Geovisualization emerged in the early 1990s in response to the increasing complexity of spatial data and advances in interactive and multimedia technologies (MacEachren, 1994). Positioned at the intersection of cartography, GIScience, cognitive science, mathematics, and scientific visualization, it develops theories, methods, and tools for the exploration, analysis, and communication of spatial information (Dykes et al., 2005; MacEachren, 1994; MacEachren & Kraak, 2001). Geovisualization combines scientific data analysis with visual reasoning, interaction, and communication to support knowledge discovery and decision-making.

Early research used interactive maps to support exploratory spatial data analysis and knowledge discovery (Dykes et al., 2005; MacEachren & Kraak, 1997). As geospatial data became more complex, geovisual analytics increasingly integrated computational models, geostatistical methods, visual interfaces, and human reasoning (G. Andrienko et al., 2007, 2011; Keim et al., 2008).

More recent research has examined how interface design, visual composition, and narrative structure shape users' interpretation, engagement, and trust (Zuo et al., 2020; Liu et al., 2025; Roth, 2021). These developments reflect a shift from treating geovisualization as a technical means of displaying spatial data toward understanding it as a human-centred process involving cognition, interaction, interpretation, and evaluation.

This broader disciplinary foundation is necessary for examining the implications of GenAI beyond isolated technical functions. It also informs the four analytical domains used in this study, *Data, Ideation, Prototyping, and Iteration,* which represent recurring areas of geovisualization practice rather than an exhaustive or sequential workflow.

### *2.2 Artistic Human-AI Collaboration*

Recent advances in GenAI have introduced new forms of human–AI collaboration in creative practice, challenging conventional distinctions between tool use, authorship, and artistic agency. In art and design, AI has been conceptualized not only as an instrument but also as a collaborator that can contribute to ideation, aesthetic exploration, and iterative refinement (Frich et al., 2019; McCormack et al., 2019).
Image-generation systems further allow users to influence colour, composition, layout, and style, creating varying degrees of shared control between human intention and algorithmic generation (Bansal et al., 2024; B. Li et al., 2019; Messer, 2024; Xiang, 2023; Zakraoui et al., 2021).

Similar forms of collaboration are emerging in geovisualization. GenAI can support the development of map concepts, provide visual references, and facilitate the iterative exploration of cartographic styles and design alternatives. Visual inputs such as artworks, existing maps, and other reference materials can also guide the generation and refinement of geovisual designs (Wang et al., 2025). Related work on map style transfer and cartographic pictogram generation further illustrates how AI-assisted methods can expand the range of available visual expressions (Christophe et al., 2022; Drews et al., 2025; Dunkel et al., 2024; Kang et al., 2019).

These developments suggest that GenAI affects not only the efficiency of visual production but also the distribution of initiative, control, and judgment between human practitioners and computational systems. However, existing studies have largely focused on individual creative tasks and system capabilities. They offer limited insight into how such collaboration affects professional judgment or interacts with the data-intensive, evaluative, and responsibility-related aspects of geovisualization.

### *2.3 GenAI Applications in Geovisualization*

GenAI applications in geovisualization span three recurring areas: geospatial data access, processing, and enrichment; generation and prototyping; and evaluation and refinement.

First, large language models (LLMs), including LLM-Geo, GeoGPT, and MapGPT, can support geospatial data access and processing by integrating multiple tools and assisting with task planning, data retrieval, database querying, filtering, and preliminary analysis (Fernandez & Dube, 2023; Feng et al., 2023a; Z. Li & Ning, 2023; Y. Zhang et al., 2023 ). GenAI has also been used for geocoding unstructured text and semantically enriching existing geospatial datasets (Chen et al., 2025; Hu et al., 2023; Juhász et al., 2023; Kim & Lee, 2024).

Second, generative methods support the production and prototyping of geospatial data and visual representations. Earlier GAN-based approaches established technical foundations for tasks such as building-footprint completion, base-map synthesis, and remote-sensing enhancement (J. Li et al., 2020; Shi et al., 2019; Wu & Biljecki, 2022; Zhu et al., 2020 ). More recent LLM- and multimodal-model-based systems further extend these capabilities through natural-language interaction, code generation, function calling, and rapid map development (Feng et al., 2023; Ning et al., 2025; Staniek et al., 2024; Wang et al., 2025; Zhang et al., 2023).

Third, GenAI is increasingly being explored for the evaluation and refinement of geovisualizations. Recent studies suggest that LLMs and multimodal models can identify cartographic principles, comment on visual hierarchy and symbolization, and provide suggestions for improving map design (Memduhoğlu, 2025; Wang et al., 2025).

Taken together, these studies address recurring activities of data handling, ideation, prototyping, and iteration, which inform the four analytical domains used in this study. However, the literature remains fragmented across scientific, creative, and technical perspectives and predominantly examines individual systems, tasks, or outputs. Recent work has also begun to raise concerns about provenance, spatial reliability, human oversight, and accountability in AI-assisted geovisualization, yet these issues are often treated as separate technical or ethical challenges rather than as interconnected questions of how responsibility is distributed across the broader workflow. As a result, existing research provides a limited understanding of how current challenges, emerging forms of human–AI collaboration, professional change, and responsibility collectively shape the development of geovisualization. Expert interviews are well-suited to addressing this gap because they can capture tacit professional knowledge, cross-activity experience, and forward-looking disciplinary perspectives that are not readily observable through evaluations of individual systems.

## 3. Methodology

We conducted semi-structured interviews with geovisualization experts to explore their perspectives on the impact of GenAI. Participants were recruited through purposive sampling to ensure relevant domain expertise (Rai & Thapa, 2004). Eligibility required a doctoral degree and teaching experience, peer-reviewed publications, or both in geovisualization or a closely related field. Direct experience with GenAI was not required, as the study sought broader disciplinary perspectives.

Participants were recruited through professional conferences, colleague referrals, academic networks, and direct email invitations. Among 31 experts contacted, 20 completed an online interview lasting approximately 60 minutes. Interviews were typically attended by three researchers, with one acting as the primary interviewer and the others contributing follow-up questions when appropriate. A shared semi-structured protocol ensured consistency while allowing participants to elaborate on issues relevant to their expertise. Interviews were conducted in English or Chinese, depending on participants' language preferences, to support clear and natural communication. Chinese transcripts and quotations used in the manuscript were translated into English by a bilingual researcher and reviewed by another researcher for semantic accuracy.

The sample size was consistent with prior qualitative research on expert and remote interviews (Caine, 2016; Hennink & Kaiser, 2022). The study aimed to capture diverse expert perspectives rather than achieve statistical representativeness.

The study received ethical approval from the university. All interviews were audio-recorded with informed consent. Participants were informed about the study purpose, recording procedures, anonymized reporting, and data protection.

### *3.1 Participants*

The 20 participants were assigned identifiers P1–P20 to support consistent attribution of anonymized quotations. Years since PhD completion were used as a consistent indicator of career stage rather than as a direct measure of professional experience. Six participants had completed their PhD within the previous five years, seven between five and fifteen years earlier, and seven more than fifteen years earlier.

The participants were affiliated with universities and research institutions in Europe, the United States, and China and represented different career stages and areas of expertise within geovisualization and related fields. They also varied in their familiarity with GenAI. Some had directly used GenAI in research, teaching, or geovisualization-related tasks, whereas others drew primarily on team-level experience or informed observation of emerging applications. This variation enabled the study to capture perspectives across different levels of professional seniority and GenAI familiarity. The sample was designed to capture diverse expert perspectives rather than to achieve statistical or institutional representativeness.

Table 1. Career stage of the 20 geovisualization experts, measured by years since PhD completion

| Years since PhD completion | Domain Experts |
|---|---|
| < 5 years | P1, P4, P5, P15, P16, P19 |
| 5-15 years | P3, P6, P9, P12, P14, P18, P20 |
| >15 years | P2, P7, P8, P10, P11, P13, P17 |

### *3.2 Interview questions and procedure*

The semi-structured interview protocol was organized into five parts, each aligned with a research question. The four analytical domains, *Data, Ideation, Prototyping, and Iteration,* were used as prompts to support discussion of recurring areas of geovisualization practice, rather than as a fixed or exhaustive workflow. Participants were encouraged to discuss overlaps between these domains and to raise issues beyond them. All participants were asked the same core questions, while follow-up questions were adapted to their expertise and responses. The full interview protocol is provided in Appendix A. Each interview lasted approximately 60 minutes.

Table 2. Structure of the semi-structured interviews.

| | |
|---|---|
| **Step 1** | Introduction to the study, informed consent, and procedures for data use and protection. |
| **Step 2** | Participants' professional backgrounds, research interests, familiarity with GenAI, and perceived current challenges in geovisualization. |
| **Step 3** | Current and anticipated impacts of GenAI across the four analytical domains, including potential applications and changes in human–AI collaboration. |
| **Step 4** | Risks associated with GenAI, activities requiring human judgment, and implications for professional practice. |
| **Step 5** | Broader reflections on future developments in geovisualization and concluding comments. |

### 3.3 Data Analysis

We conducted a combined deductive and inductive thematic analysis of the interview transcripts. All interview recordings were transcribed and anonymized before analysis. The three research questions and the four analytical domains, Data, Ideation, Prototyping, and Iteration, provided an initial deductive structure, while inductive coding was used to identify themes that extended beyond this framework.

Three researchers participated in the analysis. One researcher conducted the primary coding. The primary coder first familiarized herself/himself with the transcripts and generated initial codes, while the other two researchers subsequently reviewed the coding framework, category boundaries, thematic structure, and interpretation of the findings. Related codes were then discussed among the three researchers and grouped into broader categories and candidate themes. The coding framework was iteratively refined as new concepts emerged from the data, allowing cross-cutting themes such as human–AI collaboration, professional change, risk, and future development to be incorporated into the analysis. Differences in interpretation were resolved through discussion, and the final themes were reviewed against the interview data before being finalized.

The analysis focused on both recurring patterns and contrasting perspectives across participants. Where participant counts are reported, each participant was counted once per theme, and these counts are used descriptively rather than for statistical generalization. The overall analysis process is summarized in Figure 2.

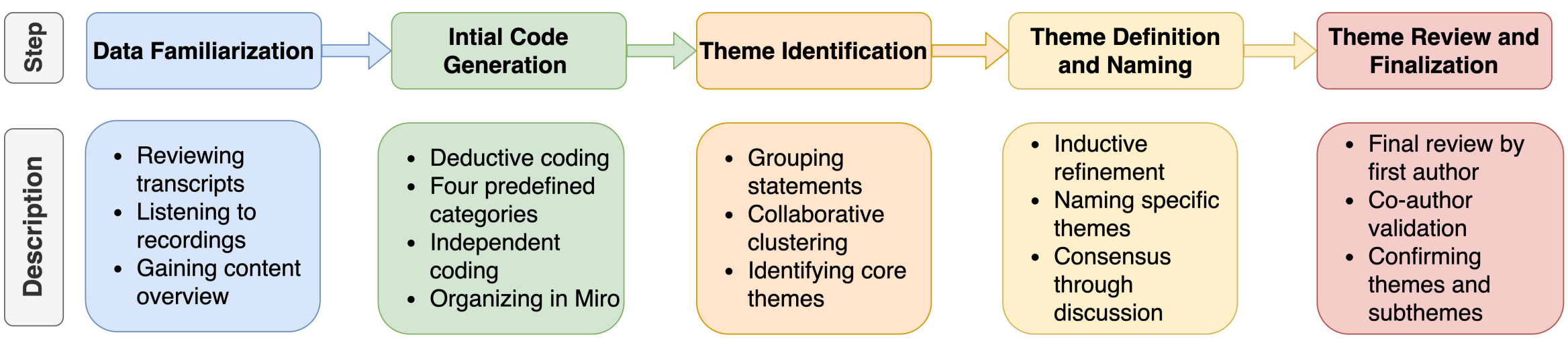

Figure 2. Overview of the deductive and inductive thematic analysis process.

## 4. Findings

Figure 3. summarizes the three themes aligned with the research questions: current challenges, impact of GenAI on geovisualization and risks from GenAI on geovisualization. This section outlines GenAI's opportunities and challenges to geovisualization. Section 4.1 addresses the current challenges within the field. Section 4.2 examines the impact of GenAI on geovisualization. Section 4.3 discusses the risks posed by GenAI. Section 4.4 summarizes the future of geovisualization based on the preceding analysis.

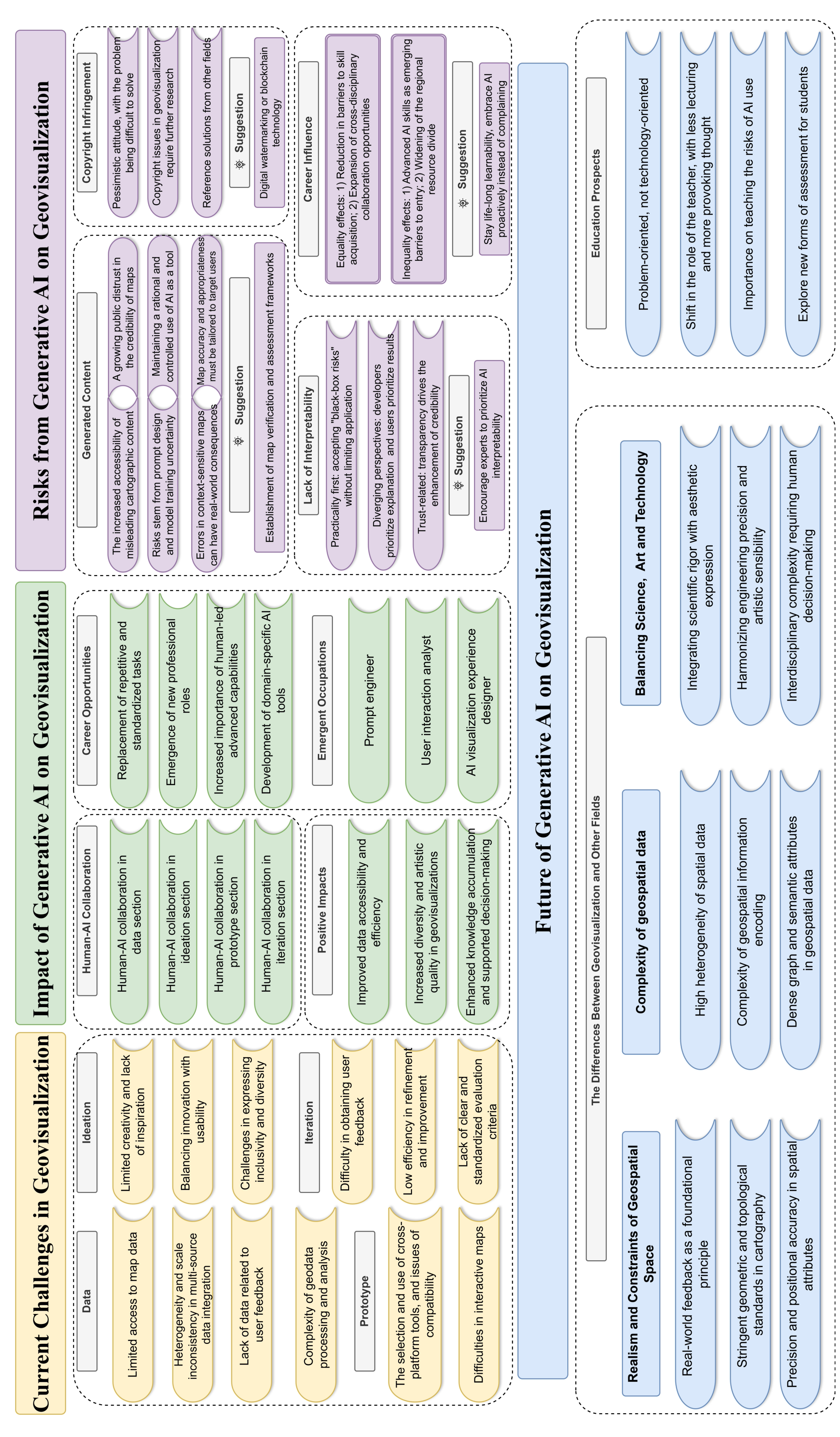

Figure 3. Overview of the investigated topics.

### *4.1 Current Challenges in Geovisualization*

Across the four analytical domains, participants described challenges throughout geovisualization practice, with particularly persistent constraints in data-related work and evaluation. More than 60% of participants identified difficulties in data acquisition, availability, integration, and processing. Limited access, inconsistent formats, heterogeneous sources, and varying levels of spatial accuracy were repeatedly mentioned. As P5 noted, "*many data sources are either not open or lack a standardized format.*" P9 similarly highlighted the scarcity of data capturing how users interact with different types of maps.

Challenges in ideation were less technical but concerned the difficulty of producing geovisualizations that are simultaneously creative, inclusive, and practically feasible. Participants described tensions between innovation, aesthetic quality, interdisciplinary requirements, and implementation constraints. As P6 explained, "*Ideation and design involve many variables, with different fields having different priorities.*" These concerns suggest that ideation depends on balancing multiple forms of professional judgment rather than simply generating new visual concepts.

Prototyping challenges centered primarily on technical implementation, particularly tool compatibility and the development of interactive and cross-platform systems. P6 identified "cross-platform tools and compatibility issues" as a major difficulty, while other participants pointed to the complexity of translating design concepts into functioning interactive systems.

In the iteration domain, participants emphasized the difficulty of obtaining meaningful user feedback and the lack of robust evaluation criteria. P12 highlighted the challenge of understanding "*how users interpret, interact with, and perceive spatial information,*" while participants also noted that highly dynamic and interactive systems remain difficult to evaluate systematically.

Taken together, these findings show that the challenges facing geovisualization extend across interconnected activities, from accessing and processing data to evaluating how spatial representations are understood and used. These existing constraints provide an important context for understanding where experts perceive GenAI as offering support and where new dependencies and risks may emerge.

### *4.2 Impact of GenAI on Geovisualization*

Participants generally viewed GenAI as an augmentative rather than a substitutive technology. Across the four analytical domains, it was perceived to expand capabilities in data handling, ideation, prototyping, and iteration, while human expertise remained important for contextual judgment, verification, and responsible decision-making (Figure 4).

#### *4.2.1 GenAI expands capabilities across geovisualization practice*

The most frequently reported benefit, mentioned by 16 of 20 participants, was improved accessibility and efficiency in data-related work. Participants highlighted applications in data search, integration, enrichment, completion, and generation. GenAI was also perceived as supporting creative exploration during ideation and accelerating map production and prototype development. Although fewer participants discussed iteration, some identified emerging uses in understanding user feedback, evaluating outputs, and supporting refinement.

More broadly, participants often described GenAI as an assistant that extends existing capabilities rather than independently completing geovisualization work. Its perceived strengths were concentrated in information retrieval, content generation, and rapid exploration, particularly where tasks could be more readily specified or standardized.

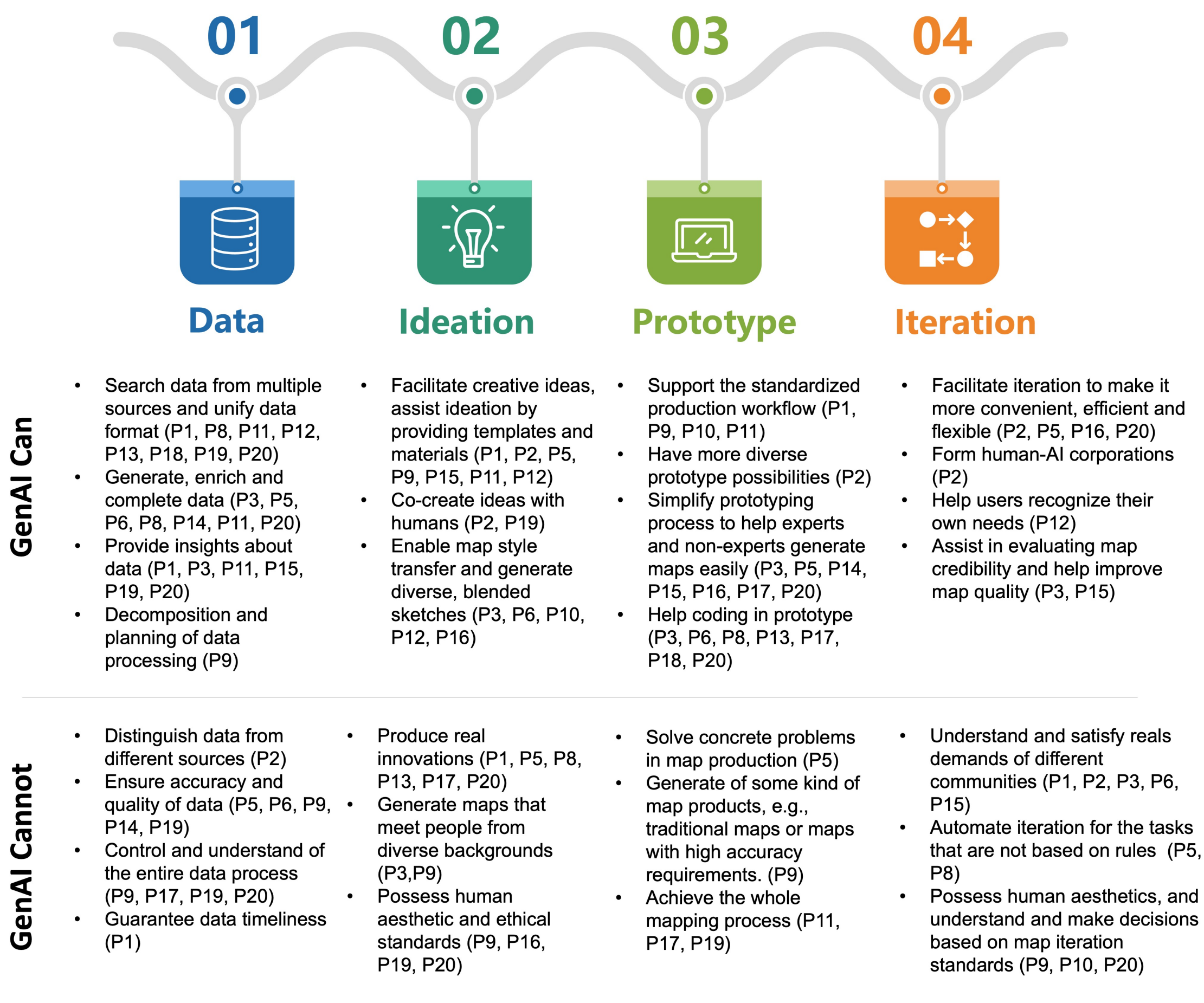

Figure 4. Activities that can and cannot be completed by GenAI in the geovisualization workflow.

4.2.2 *Human expertise remains central to judgment and verification*

Participants were more cautious about activities requiring creativity, aesthetic judgment, contextual understanding, and ethical decision-making. Several questioned whether current systems could independently assess the appropriateness of a visualization or respond adequately to diverse user needs. As P15 noted, "*Ordinary ideas can be realized, but true innovation still requires human involvement.*"

Spatial accuracy was another recurring concern. Participants emphasized that visually plausible outputs may still contain errors in map content or location and, therefore, require expert review. As P9 observed, "*Current generative models may introduce inaccuracies in specific map content or locations*." Human involvement was therefore perceived as particularly important for validating spatial information, adapting outputs to specific users and contexts, and making final decisions about their use.

4.2.3 *Professional roles shift toward higher-order judgment*

These changes in the division of labor between humans and GenAI were also reflected in participants' expectations for professional practice. Routine and standardized tasks, including parts of data processing, rendering, and basic design, were expected to become increasingly automated. Consequently, skills primarily centered on software operation were perceived as less distinctive.

At the same time, participants anticipated increasing importance for capabilities related to AI interaction, workflow design, evaluation, and domain-specific judgment. P5 summarized this shift by noting that "*tool usage will become secondary; what matters more is learning how to interact with the existing workflow.*" Rather than eliminating professional expertise, participants therefore expected GenAI to reposition its value toward directing, evaluating, and taking responsibility for AI-assisted geovisualization.

## *4.3 Risks of GenAI for Geovisualization*

### *4.3.1 Reliability and Appropriateness of Generated Content*

Participants were particularly concerned that GenAI could make misleading or inaccurate geovisualizations easier to produce and harder to distinguish from reliable outputs. As P11 warned, "*it will become easier to generate false information, so people must be more careful.*" Bias in training data and prompting was also seen as a potential source of distortion, while inappropriate outputs could be especially consequential in sensitive application contexts. These concerns led many participants to emphasize the need for stronger verification and human review before AI-generated geovisualizations are used or communicated.

### *4.3.2 Provenance, Ownership, and Interpretability*

Participants also raised concerns about whether AI-generated outputs could be traced, attributed, and understood. Copyright and ownership were regarded as difficult to resolve because the sources of training data and individual visual elements may be unclear. P3 observed that "*copyright infringement has always existed, and the problem may become even more pronounced in the current era.*" Some participants proposed mechanisms such as watermarking and provenance tracking, although these were discussed as possible rather than established solutions.

Interpretability concerns were closely linked to trust. Participants noted that the need to understand model processes depends on the intended use: developers may require greater transparency, whereas some end users may focus primarily on whether outputs meet their needs. Overall, participants viewed traceability and interpretability as important in determining when AI-assisted geovisualizations can be trusted and used responsibly.

#### *4.3.3 Uneven Professional and Social Consequences*

Participants expressed contrasting views on the broader consequences of GenAI for access and professional development. Some believed that AI assistance could lower technical barriers, make basic geovisualization more accessible to non-specialists, and support interdisciplinary collaboration. Others cautioned that new competencies in AI interaction and unequal access to computational resources could create new professional and regional inequalities.

Rather than simply removing barriers, participants therefore perceived GenAI as potentially redistributing them. This shift increases the importance of continued learning and raises broader questions about who has access to emerging tools, skills, and opportunities in AI-assisted geovisualization.

### ***4.4 Domain-Specific Implications for Geovisualization***

Participants identified several characteristics that distinguish geovisualization from other domains of GenAI application. First, geovisualization is constrained by geographic reality. Maps and spatial representations must preserve locational accuracy and spatial relationships, including geometry, topology, and scale. As P1 and P4 emphasized, geovisualization "*reflects the real world and cannot be detached from reality.*" This makes visually plausible but spatially incorrect outputs particularly problematic.

Second, participants highlighted the complexity of geospatial data. Unlike general text or image content, geospatial information combines geometric, topological, semantic, and relational structures. P12 noted that geospatial information "*is not just an image, and it is not text,*" but involves multiple forms of spatial encoding. Participants questioned whether general-purpose models alone are sufficient for tasks that require robust spatial reasoning.

Third, participants described geovisualization as a practice that must balance scientific accuracy, visual expression, and technical implementation. P9 summarized this tension by noting that geovisualization combines "*engineering precision with artistic sensibility.*" This combination places particular importance on human judgment, especially when generated outputs must remain both spatially valid and visually meaningful.

These characteristics also shaped participants' views on education. As GenAI lowers some technical barriers, participants argued that geovisualization education should place greater emphasis on problem framing, critical evaluation, spatial reasoning, and ethical judgment rather than tool operation alone. Several also called for new approaches to assessment and greater attention to AI-related risks, including data ethics, fairness, and responsibility. In this sense, the growing use of GenAI was seen not as reducing the need for expertise, but as changing what expertise in geovisualization should involve.

# 5. Discussion

## *5.1 GenAI shifts the bottleneck from production to judgment*

Our findings suggest that GenAI does not eliminate the bottlenecks of geovisualization; rather, it relocates them. Traditional geovisualization has long faced constraints related to data availability, technical implementation, and the production of visual representations. GenAI can reduce some of these barriers by accelerating data processing, map generation, and visual exploration. However, as production becomes easier, the more consequential difficulties increasingly concern whether generated outputs are spatially valid, contextually appropriate, and sufficiently reliable for their intended use.

This shift is particularly important because longstanding risks in geovisualization, including misleading representation and copyright infringement, do not disappear with automation (Belyea, 2018; Harris, 1998; Rodriguez, 2019). Instead, GenAI can increase the speed and scale at which realistic spatial data, maps, and visual narratives are produced (Kang, 2020; Sun et al., 2020; Wei et al., 2025). The resulting challenge is not simply that inaccurate or misleading content can be generated, but that visually plausible outputs may obscure spatial errors, making them harder to recognize. In this context, verification becomes a more central bottleneck: practitioners must assess not only whether an output appears convincing, but whether its locations, spatial relationships, semantics, and representational choices are valid.

A similar shift occurs in relation to provenance and interpretability. The opacity of training data and generative processes complicates the attribution of authorship, ownership, and responsibility (Samuelson, 2023). Computational provenance can help document data sources and transformations (Freire et al., 2008), but traceability alone cannot determine whether an output is accurate, fair, or appropriate. Likewise, limited interpretability may make it difficult to identify the sources of spatial errors or embedded biases. The bottleneck extends beyond technical verification to questions of judgment: practitioners must decide what level of uncertainty is acceptable, what evidence is sufficient for validation, and whether an AI-generated geovisualization should be used at all.

These changes also make responsibility more consequential. When GenAI contributes to data handling, design, generation, and refinement, the final output may no longer be attributable to a single transparent sequence of human decisions. This does not remove human responsibility; instead, it increases the need to make responsibility explicit at critical points of validation and use. The central challenge is therefore not merely how to produce geovisualizations more efficiently, but how to ensure that expanded generative capacity is matched by sufficient professional judgment, verification mechanisms, and accountable decision-making.

### *5.2 Human expertise is being redefined, not displaced*

Our findings suggest that GenAI is changing not only how geovisualizations are produced, but also what constitutes professional expertise in the field. GenAI can improve efficiency and support creative exploration (Gambacorta et al., 2024; Zhou & Lee, 2024), yet these capabilities do not remove the need for expert judgment. Instead, as routine production and tool operation become increasingly assisted, professional value shifts toward abilities that are more difficult to formalize, including spatial reasoning, contextual interpretation, aesthetic evaluation, and ethical judgment.

This shift can be partly understood through the tacit nature of expert knowledge. Much professional knowledge cannot be fully expressed as explicit rules, a point captured by Polanyi's paradox that "we know more than we can tell" (Autor, 2014). Creative and professional judgment often draws on experience, intuition, perception, and context rather than solely on codified procedures (Runco & Chand, 1995). In geovisualization, this is particularly important because valid representation requires more than generating visually appealing outputs. Practitioners must assess whether spatial relationships are plausible, whether a design communicates appropriately to particular users, and whether aesthetic choices remain consistent with geographic meaning. GenAI may accelerate the generation process, but these situated forms of judgment remain closely tied to domain expertise (N. Andrienko & Andrienko, 2025; Hitsuwari et al., 2023).

This also changes the role of creativity in human–AI collaboration. Rather than understanding creativity only as the production of novel outputs, GenAI makes selection, adaptation, rejection, and contextualization increasingly important parts of creative work. AI-generated alternatives can stimulate ideation and reduce the burden of starting from a blank page (Kaufman & Sternberg, 2006; Kenower, 2020; Preshaw, 2025), while supporting exploratory processes such as early-stage ideation and style transformation (Kang et al., 2024; Tao & Xu, 2023). Such assistance may reduce routine cognitive effort and free attention for higher-level reasoning and decision-making (Lubart, 2005; Shneiderman, 2022). However, the value of this support depends on whether users remain actively involved in evaluating and shaping the outputs rather than accepting generated alternatives by default.

Therefore, the core idea is not simply to preserve human creativity while letting artificial intelligence handle routine tasks. Rather, the very meaning of expertise is being redefined. As technical execution becomes less distinctive, geovisualization professionals may be increasingly valued for their ability to frame problems, reason spatially, interpret context, evaluate visual and ethical trade-offs, and decide when AI-generated outputs are appropriate for use. GenAI thus shifts expertise away from tool proficiency alone and toward domain judgment: the capacity to understand not only how to produce a geovisualization, but also why a particular representation should be produced, whether it is valid, and what consequences may follow from its use.

### *5.3 Geovisualization requires a domain-specific form of responsible AI*

Our findings suggest that responsible AI in geovisualization cannot be reduced to generic concerns about model transparency, bias, or human oversight. Geovisualization occupies a distinctive position between scientific analysis, visual expression, and technological implementation (Çöltekin et al., 2018; Dykes et al., 2005). GenAI can augment all three dimensions by supporting spatial analysis and pattern discovery (Klemmer et al., 2019), expanding visual styles (Wang et al., 2025), and accelerating the production of maps and spatial scenes (Ganguli et al., 2019). However, the same capabilities create tensions that are specific to spatial representation: outputs must remain geographically valid while also being visually meaningful and technically usable.

The first requirement is therefore spatial and epistemic validity. Geovisualization is expected to preserve relationships among location, geometry, topology, scale, and spatial semantics. GenAI complicates this expectation because visually plausible outputs may still be spatially incorrect or difficult to verify (Kang et al., 2023; Lin & Zhao, 2025; Wang et al., 2025). The opacity of generative models further complicates interpretability, reproducibility, and confidence in the production of spatial representations (Lipton, 2018). Responsible use requires more than visually convincing outputs. Provenance mechanisms, uncertainty communication, and domain-specific verification are needed to make data sources, transformations, model interventions, and output limitations more visible (Cordes et al., 2020; Werder et al., 2022).

The second requirement is responsible mediation between accuracy and expression. Geovisualization is not a form of unconstrained image generation: creative and aesthetic choices remain bounded by geographic meaning and the consequences of representation (MacEachren & Kraak, 2001). GenAI can lower barriers to visual exploration and broaden participation in map-making, but increased accessibility also makes it easier to produce persuasive representations without sufficient cartographic or spatial expertise. The central issue is therefore not whether GenAI should enhance creativity, but how to introduce generative flexibility without weakening spatial credibility. Future systems should support users in identifying uncertainty, checking spatial consistency, and distinguishing exploratory or synthetic outputs from representations intended to function as evidence.

The third requirement concerns the distribution of responsibility across the geovisualization process. In conventional workflows, errors can often be associated with identifiable decisions in data preparation, analysis, or cartographic design. GenAI introduces additional layers of dependency through training data, model behavior, prompts, automated transformations, and iterative human–AI interactions. Risks may therefore emerge and propagate across multiple stages rather than remain isolated within a single operation (Choi et al., 2021; Fulman et al., 2024; Marasinghe et al., 2024). Responsibility therefore needs to be considered across model providers, system developers, geovisualization practitioners, institutions, and users. Provenance can help document these contributions (Freire et al., 2008), but responsible practice also requires clarity about who validates outputs, who authorizes their use, and who remains accountable when spatial representations inform consequential decisions.

This perspective reframes the relationship among science, art, and technology in AI-assisted geovisualization. GenAI does not simply strengthen each dimension independently; it increases their interdependence. Greater technological capacity expands opportunities for analysis and expression, but also raises the demands placed on scientific verification, professional judgment, and governance. A domain-specific responsible AI framework for geovisualization should therefore integrate spatial validity, provenance, uncertainty communication, human oversight, and accountable use across the full human–AI workflow. The challenge for the field is not merely to balance science, art, and technology, but to ensure that advances in generative capability are matched by equally robust mechanisms for maintaining the credibility and responsibility of spatial knowledge.

### *5.4 Implications, Limitations, and Future Directions*

The shifts identified in this study have implications for professional practice, education, and system design. As GenAI becomes more deeply embedded in geovisualization, practitioners will increasingly need to supervise, validate, and contextualize AI-generated outputs rather than focus primarily on manual production. Professional expertise will therefore depend not only on technical proficiency, but also on spatial reasoning, critical evaluation, contextual judgment, and accountable decision-making.

Education should evolve accordingly. Training in geovisualization should move beyond software proficiency to strengthen spatial reasoning, critical evaluation, ethical judgment, and literacy in AI-assisted workflows. Future systems should likewise support responsible human–AI collaboration by making provenance and uncertainty more visible, enabling spatial validation and human review, and maintaining traceable decision processes. These directions suggest that the future of geovisualization will depend less on maximizing automation than on ensuring that increasing generative capacity is matched by appropriate forms of expertise, verification, and accountability.

These implications should be interpreted in light of the study's scope. The four analytical domains—Data, Ideation, Prototyping, and Iteration—were used to structure the inquiry rather than to represent an exhaustive or sequential model of geovisualization practice. Although this framework supported comparison across interviews, it may not have explicitly captured more complex activities and interactions. Future research could therefore examine how GenAI operates across more recursive, organizational, and deployment-oriented processes.

The purposive sample of 20 experts was also intended to provide an expert-grounded synthesis rather than a representative account of the field. Participants differed in geographic context, career stage, and familiarity with GenAI, but the study did not systematically compare these differences. Future work could investigate how perspectives vary across regions, professional settings, and levels of direct GenAI experience. The flexibility of semi-structured interviews allowed participants to elaborate on issues relevant to their expertise, but also introduced variation in the depth of discussion across topics. Further studies could complement expert interviews with observational studies, longitudinal research, and evaluations of real-world AI-assisted geovisualization practices.

These limitations define the boundaries of the present study and identify a broader research agenda. Future research should examine how AI-assisted geovisualization is validated, governed, taught, and practiced across different institutional and social contexts, rather than focusing on what GenAI can generate.

## 6. Conclusion

This study drew on interviews with 20 geovisualization experts to examine how GenAI is reshaping geovisualization practice. The findings show that GenAI is not simply introducing new tools or efficiencies. Rather, it is redistributing where difficulties, expertise, and responsibility are concentrated across geovisualization.

First, as GenAI supports data handling, ideation, prototyping, and iterative refinement, key bottlenecks are increasingly shifting from production toward judgment and verification. Second, as routine technical tasks become more automated, professional value moves beyond tool proficiency toward spatial reasoning, contextual interpretation, aesthetic and ethical judgment, and the ability to evaluate when AI-generated outputs are appropriate for use. Third, as GenAI becomes embedded across multiple stages of practice, risks and responsibilities become more distributed across models, developers, practitioners, institutions, and users.

These shifts are particularly significant in geovisualization because spatial representations are constrained by geographic reality and must balance scientific validity, visual expression, and technical implementation. Responsible GenAI in this field requires more than generic concerns about transparency or bias; it also depends on spatial validation, provenance, communication of uncertainty, human oversight, and accountable use.

Overall, this study provides an expert-grounded perspective on the broader implications of GenAI for geovisualization. The future development of the field will depend not only on expanding generative capabilities, but also on strengthening the forms of judgment, expertise, and governance needed to ensure that AI-assisted spatial knowledge remains credible, meaningful, and responsibly used.

**Disclosure statement**

No potential conflict of interest was reported by the authors.